\documentclass[epj]{svjour}
\usepackage{graphics}
\usepackage{color}
\begin{document}
\title{Emergence of self-similarity in football dynamics}
\author{Akifumi Kijima\inst{1} \and Keiko Yokoyama\inst{2,3} 
\and Hiroyuki Shima\thanks{Corresponding author: hshima@yamanashi.ac.jp}\inst{4} 
\and Yuji Yamamoto\inst{5}
}                     
%
%
\institute{\inst{1} Graduate School of Education, University of Yamanashi, 4-4-37, Takeda, Kofu, Yamanashi 400-8510 Japan\\
  \inst{2} Division of Applied Physics, Faculty of Engineering, Hokkaido University, Sapporo, Hokkaido 060-8628 Japan\\
  \inst{3} Japan Society for the Promotion of Science, 5-3-1, Koujimachi, Chiyoda-ku, Tokyo 102-0083 Japan\\
  \inst{4} Department of Environmental Sciences \& Interdisciplinary Graduate School of Medicine and Engineering, University of Yamanashi, 4-4-37, Takeda, Kofu, Yamanashi 400-8510 Japan\\
  \inst{5} Research Center of Health, Physical Fitness and Sports \& 
Department of Psychology and Human Developmental Sciences, Nagoya University, Chikusa, Nagoya 464-8601 Japan}
\date{Received: date / Revised version: date}
%
\abstract{
The multiplayer dynamics of a football game is analyzed to unveil self-similarities in the time evolution of player and ball positioning. Temporal fluctuations in both the team-turf boundary and the ball location are uncovered to follow the rules of fractional Brownian motion with a Hurst exponent of $H \sim 0.7$. The persistence time below which self-similarity holds is found to be several tens of seconds, implying a characteristic time scale that governs far-from-equilibrium motion on a playing field.
} 
\maketitle
%


\section{Introduction}

Association football (also known as `soccer') is
one of the most popular sports in the world.
Its popularity, which cuts across different cultures and borders,
can be accounted for by two phrases: simple requirements and complex dynamics.
In contrast to other popular ball sports, such as hockey and baseball, players need little equipment besides the  ball.
For the audience, on the other hand,
the enjoyability of football arises from the complexity of the multiplayer dynamics
displayed on the field
stemming from the high number of players actively interacting with each other
and the almost uninterrupted flow of the ball \cite{MendesEPJB2007}.
During a match, players adjust their positions in accordance with the interactions of
other teammates and opponents as well as their distance from the ball and goal.
As the set of preferred positions at any given instant may alter drastically in the next moment
owing to sudden movement of the ball or tactical management \cite{GrunzHumMovSci2012}, the local and global configurations of players vary constantly.
The fact that so many different strategies and configurations are possible means that
a football match never gets boring.

In academic terminology, the dynamics of football players is said to
result from a succession of symmetry-breaking performances \cite{YokoyamaPlos2011,YamamotoPlos2011}.
From a local standpoint, one attacker carrying the ball
will randomly break local symmetry with his/her immediate defenders
in order to successfully pass or dribble against them.
In a global sense, however, such individual actions are not completely random,
but should  correlate and harmonize toward a common aim
({\it i.e.} team scoring),
in a way that is analogous to many stochastic processes observed in social phenomena \cite{CastellanoRMP2009}
or in cell biology \cite{SWangPNAS2011}.
These facts motivated us to consider whether
the accumulation of individual decisions in a football game
induces the spontaneous formation of patterns,
{\it i.e.} spatiotemporal patterns of player coordination.
To explore such emergent properties,
it is necessary to complete a series of real-time analyses
covering the collective motion of all of the players as well as the ball
during the total game period.
To date, this approach garnered little attention
especially in comparison to the high level of interest in 
scoring data statistics \cite{BittnerEPL2007,BittnerEPJB2009,HeuerEPL2010},
probability analyses \cite{BenNaimJSP2013,daSilvaCPC2013},
and complex network approaches \cite{OnodyPRE2004,DuchPlosOne2010,CarlosJSSC2013}
to the game of football.

In this contribution, we demonstrate  experimental evidence
for self-similarity hidden within  complex football dynamics.
Based on a time-series analysis of multiplayer configurations
extracted from real match data,
we ascertained that the time-series variations both in the team-turf boundary
and the ball location lie within the realm of the fractional Brownian motion (fBm).
From this work, a Hurst exponent $H$
(the dimensionless estimator for self-similarity in fBm)
of $H \sim 0.7$ was found,
implying the existence of a memory effect in football dynamics.

\section{Methodology}

\begin{figure}
\resizebox{0.45\textwidth}{!}{%
  \includegraphics{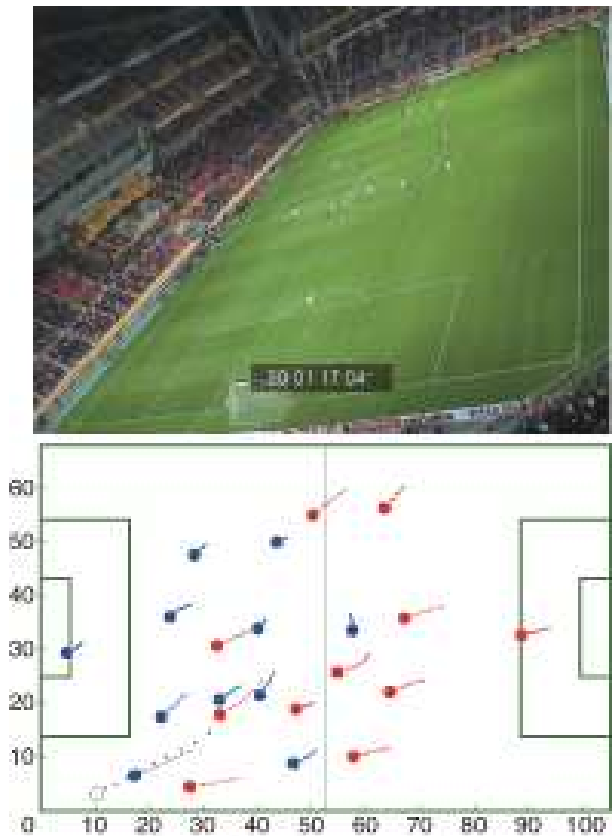}
}
\vspace*{5mm}
\caption{(color online)
Players' and ball's positions captured on video (top)
and on pitch coordinates (for 2 s: bottom).
Two-dimensional coordinates on the video screen were transformed into
105 $\times$ 68 m pitch coordinates
using a direct linear transformation technique \cite{WinterBHM2007}.
}
\label{fig.1}
\end{figure}

We elaborate upon
the particulars of two football matches:
a semifinal game in the 2008 FIFA Club World Cup
played by Gamba Osaka (the Asian champion club) and
Adelaide United (the champion of Oceania), and
a regular season game in the 2011 Japanese soccer league
featuring the Urawa Red-Diamonds versus the Yokohama 
F-Marinos\footnote{Japanese soccer league data were recorded by 
Data Stadium Inc. (Tokyo, Japan), using TRACAB$^{\rm TM}$ Player 
Tracking System (ChyronHego, Sweden).}.
Using a digital video camera, we filmed both games at 30 Hz
in order to record all of the time-varying positions of the players and ball
over the entire course of each match.
The resulting data were then transformed into two-dimensional coordinates
along the horizontal ($x$: length = 105 m)
and vertical ($y$: width = 68 m) axes.
Figure \ref{fig.1} shows an example of a snapshot of player configuration
(solid circles) based on individual short-duration paths (dashed lines: 2 s).
The ball position at this moment is depicted with an open circle.
We accumulated this coordinate data into a series running through the game
and then analyzed the player configuration dynamics (as well as the ball movement)
by focusing on two characteristic dynamic quantities explained below.
For brevity, our final analyses omit irrelevant time intervals during which
the ball left the pitch or play was interrupted
by penalties or set plays.

There are two important measures of football dynamics:
the time evolution of the ball location and that of the `frontline', {\it i.e.}
the boundary separating the respective team `turfs'.
The significance of the frontline is that
its back-and-forth fluctuations over time describe
the process by which the two teams compete for territory.
At  kick-off, this boundary coincides with the centreline of the pitch, while
its subsequent shifts and spatial distortions from this state
reflect the aggregated movement of multiple players over the pitch.

\begin{figure}
\resizebox{0.48\textwidth}{!}{%
  \includegraphics{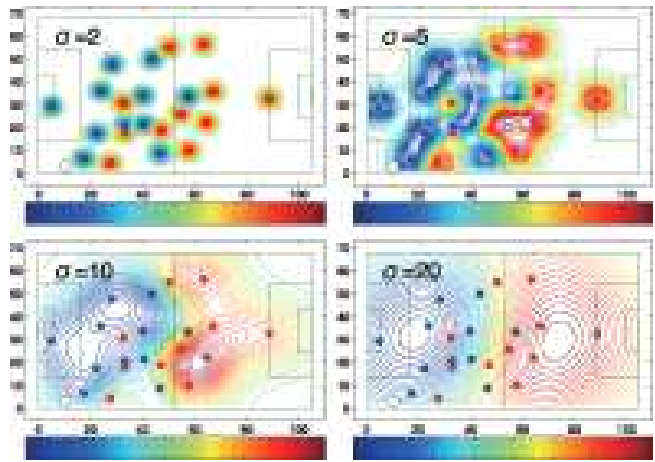}
}
\vspace*{5mm}
\caption{(color online)
Four examples of relative power distribution in the Osaka-Adelaide game.
The colour bars under the panels indicate the value of $P_{\rm A}-P_{\rm B}$,
the domination trend for each team.
Right (Red): Osaka dominates, Left (Blue): Adelaide dominates.
All panels are drawn using the same player configuration data,
while the spatial profile of each contour plot
differs owing to the different $\sigma$-value applied.}
\label{fig.2}
\end{figure}

The turf boundary is derived from a superposition of
the two-variable function $p_i(x,y)$, called the `power of domination',
which indicates the degree to which
the $i$th player at position $(x_i, y_i)$
influences the pitch position $(x,y)$.
The function $p_i(x,y)$ is defined by
\begin{equation}
p_i(x,y) = \frac{1}{2\pi\sigma}
\exp\left[ \left(\frac{x-x_i}{\sigma}\right)^2 + \left(\frac{y-y_i}{\sigma}\right)^2 \right],
\label{e1}
\end{equation}
where $\sigma$ denotes the influence distance of the individual player.
The sum of $p_i$ over the members of team A, designated by
\begin{equation}
P_{\rm A}(x,y) = \sum_{i=1}^{N} p_i(x,y) \;\; \mbox{with $N=11$},
\label{e2}
\end{equation}
determines the degree of domination of team A at position $(x,y)$;
the same definition applies to $P_{\rm B}(x,y)$.
The turf boundary can be defined as the contour line
along which $P_{\rm A}(x,y) =  P_{\rm B}(x,y)$ is satisfied.

\begin{figure}
\resizebox{0.48\textwidth}{!}{%
  \includegraphics{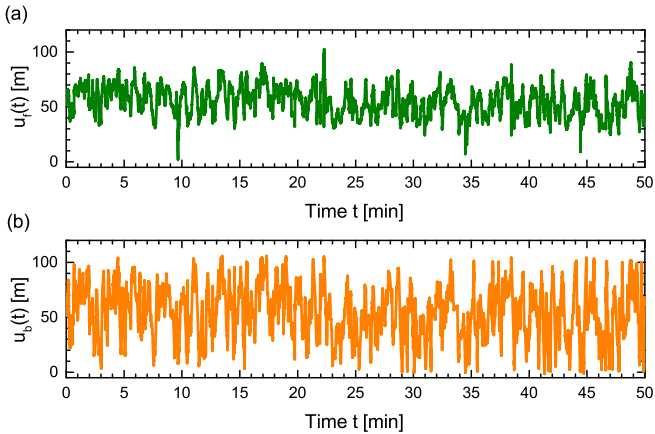}
}
\vspace*{5mm}
\caption{(color online)
Time evolution of the longitudinal position of the frontline ($u_{\rm f}$) and the ball ($u_{\rm b}$)
extracted from the Osaka-Adelaide game.}
\label{fig.3}
\end{figure}

Figure \ref{fig.2} shows a contour plot of the difference
$P_{\rm A}(x,y) - P_{\rm B}(x,y)$ on the $x$-$y$ plane
based on data evaluated from the Osaka-Adelaide game;
positive values were assigned to the Osaka turf (coloured in red) and
negative to the Adelaide turf (blue).
All four plots in Fig.~\ref{fig.2}
illustrate the relative power distribution at a single moment,
while the spatial profiles strongly depend on the parameter $\sigma$.
It is reasonable to set $\sigma = 20$ m
because in actual game play, the defenders (three or four at most) for each team
will favour evenly spacing their positions
across the width of the pitch $= 68$ m,
and the positions of the defenders will regulate the relative distances between the other players,
particularly the forwards of the opposing team.
Thus, the influence distance of each individual player will be $\sim 20$ m on average,
{\it ca.} one-third of the width of the pitch.
Based on this natural assumption, we can identify the turf boundary with a single green curve
in the contour plot (see right-bottom panel in Fig.~\ref{fig.2}) that passes across
the pitch in a nearly linear direction.

The longitudinal position of the boundary curve,
represented by $x=f(y)$ in the $x$-$y$ coordinate system,
can be quantified based on the distribution function
$P(x_0) \propto \int \delta[x_0-x(y)] dy$.
The median value of $x$ extracted from the graph of $P(x_0)$ can be defined as
the longitudinal position of the frontline expressed as a time-dependent function by $u_{\rm f}(t)$.
Similarly, the longitudinal ball position can be represented by $u_{\rm b}(t)$.

\section{Result 1: Time evolution data}

\begin{figure}
\resizebox{0.32\textwidth}{!}{%
  \includegraphics{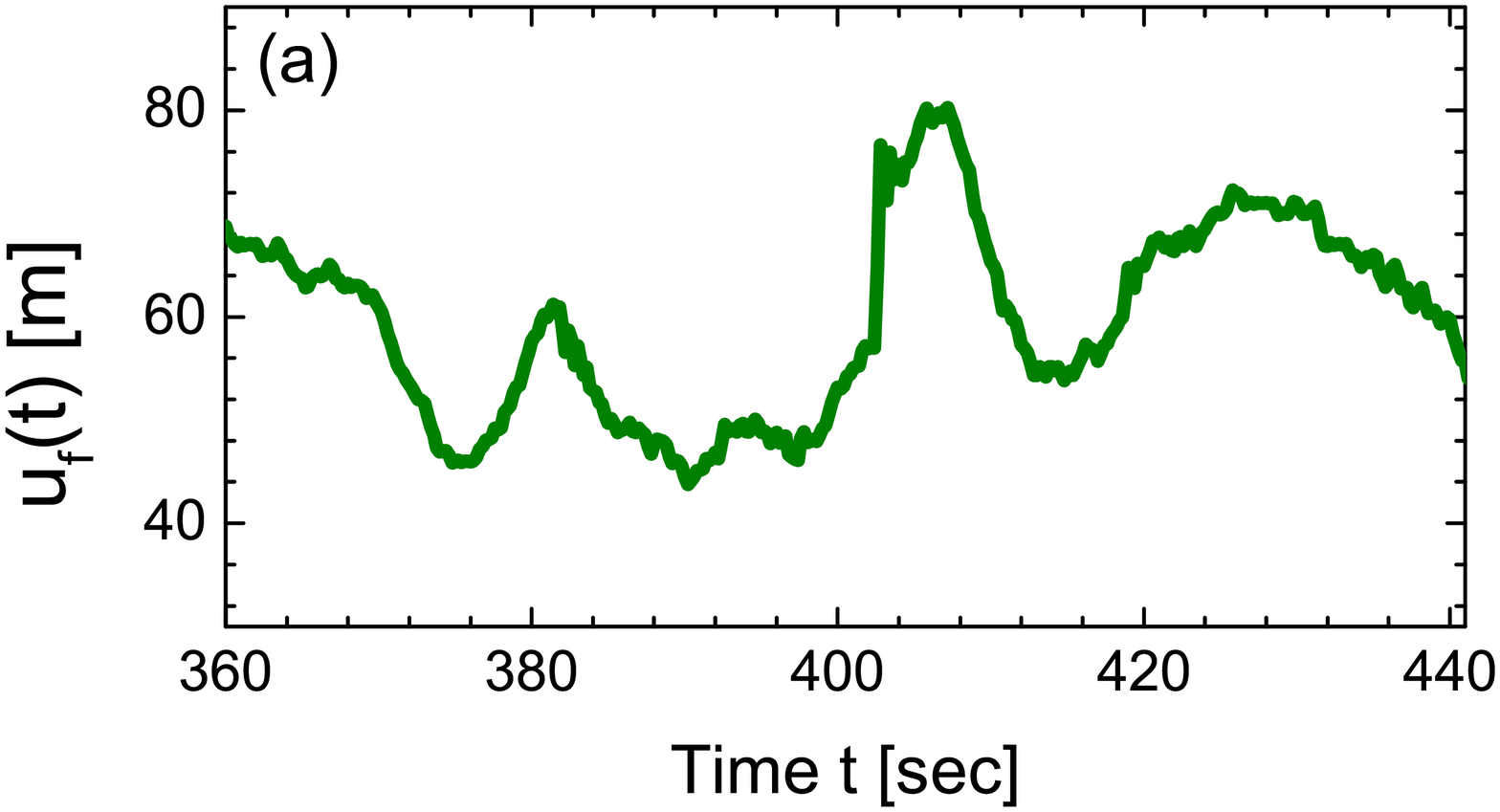}
}
\resizebox{0.32\textwidth}{!}{%
  \includegraphics{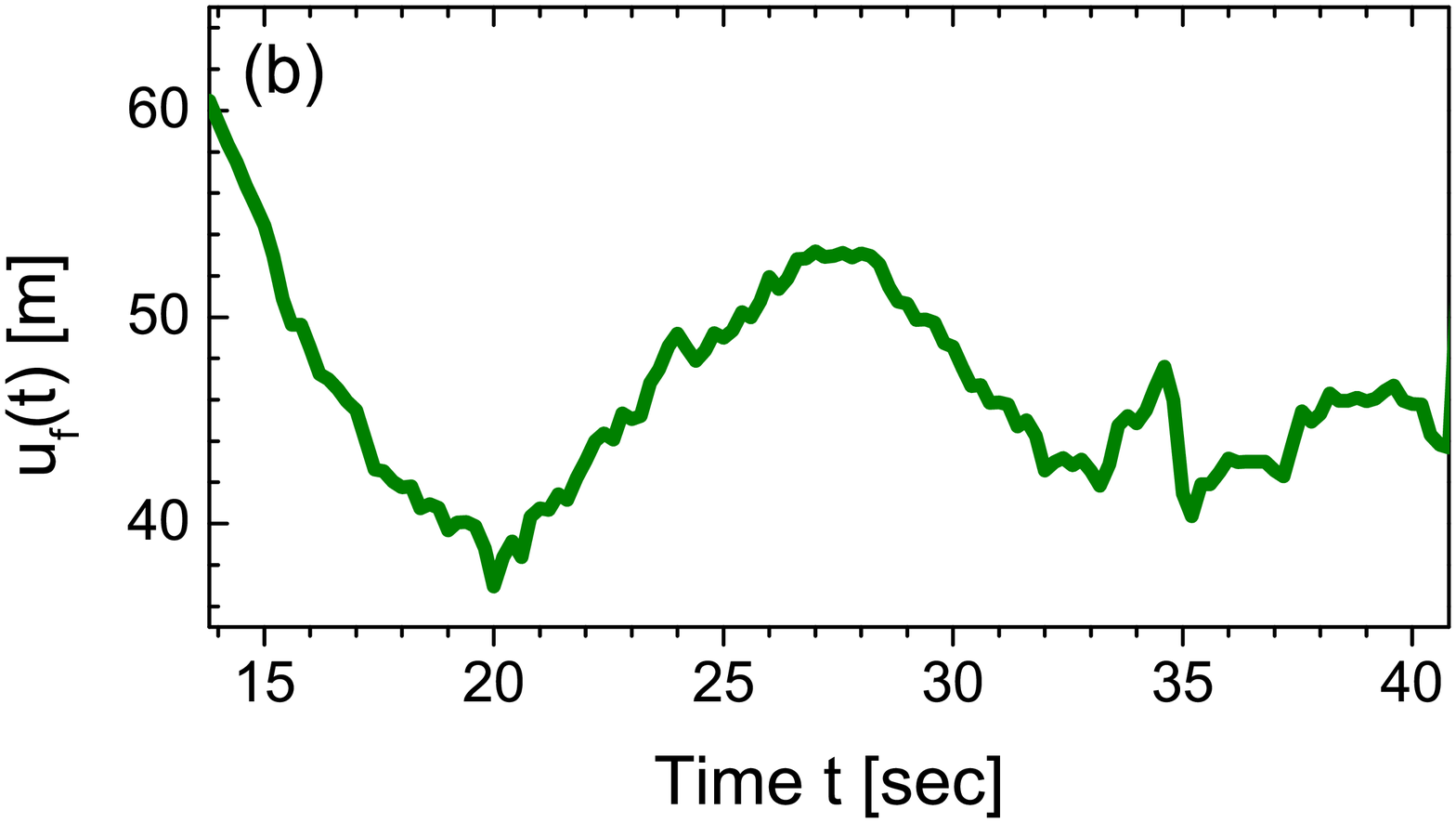}
}
\resizebox{0.32\textwidth}{!}{%
  \includegraphics{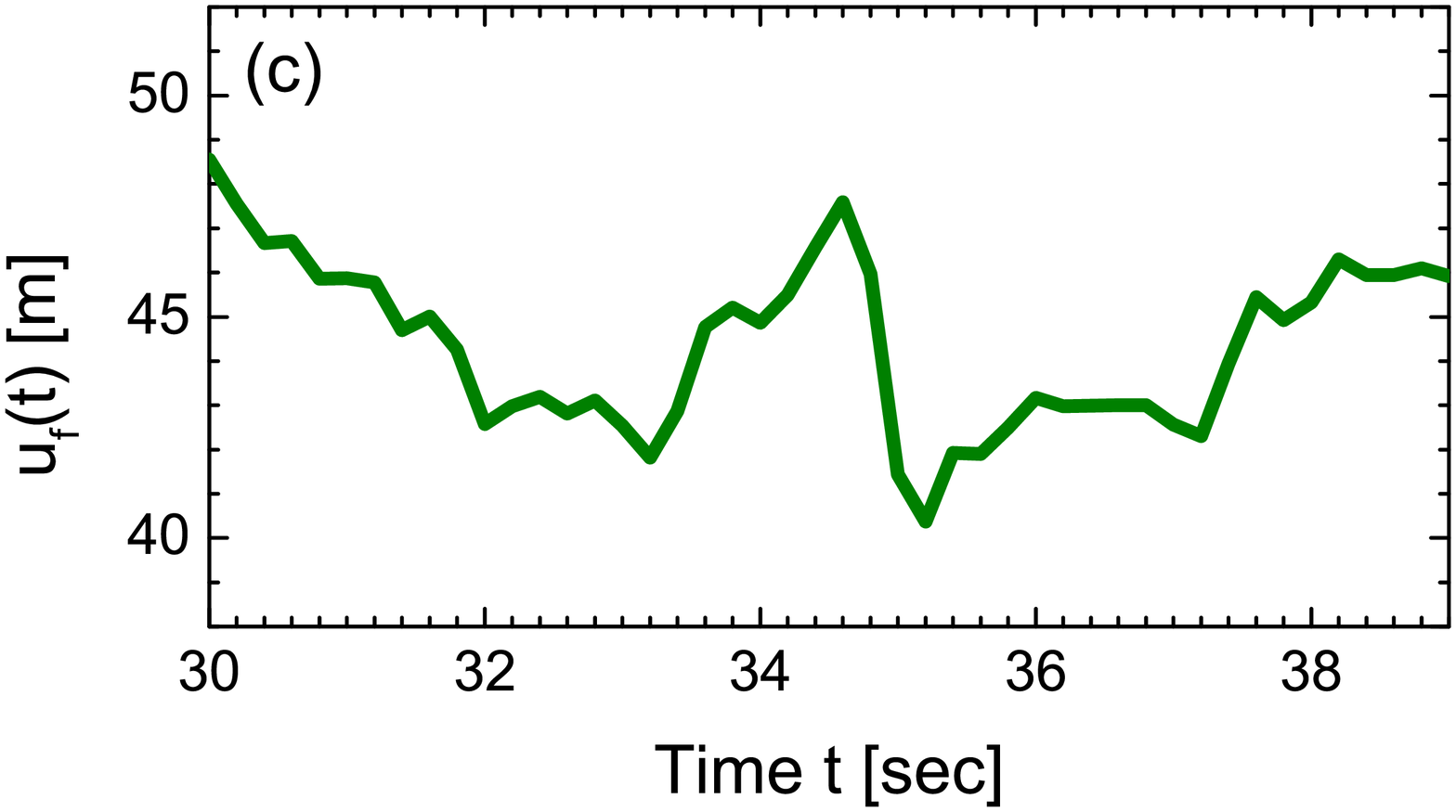}
}
\vspace*{5mm}
\caption{(color online)
Self-similarity of  frontline dynamics $u_{\rm f}(t)$.
The middle (or bottom) panel is a rescaling of the top (middle) panel
along the horizontal axis by a factor of three.
The rescaled panel has nearly the same distribution of jumps as the initial panel.}
\label{fig.4}
\end{figure}

Figure \ref{fig.3} shows the time evolution of $u_{\rm f}(t)$ and $u_{\rm b}(t)$
as obtained from  Osaka-Adelaide game data.
The sequence of the spikes here indicates intermittent chances and crunches performed near the goals
by the respective teams.
The amplitude of oscillation in $u_{\rm b}(t)$ is greater than that in $u_{\rm f}(t)$, which is expected
owing to the smaller `mass' of the ball than that of 22 moving players.

One striking discovery is that anisotropic self-similarity ({\it i.e.} self-affinity) \cite{BouchaudPhysRep1990}
can be detected in the two profiles.
We were able to show that an increment 
\begin{equation}
\Delta u_j(\tau) = u_j(t+\tau) - u_j(t) \;\; [j= \mbox{f or b}]
\end{equation}
within a time interval $\tau$
has identical probability to one
$\Delta u_j(a\tau) = a^H \cdot \Delta u_j(\tau)$
rescaled by an appropriate value of $H$.
The self-similar structure of series $u_{\rm f}(t)$, for instance,
can be seen in the sequence of three panels shown in Fig.~\ref{fig.4}.
Shrinking the time scale in successive increments of one-third from the left to the right panel
leaves the profiles seemingly unchanged.
We confirmed that such self-similar behaviour in $u_{\rm f}(t)$ and $u_{\rm b}(t)$
appeared in the data from the other game, Urawa versus Yokohama.
It is interesting to note that self-similarity in football dynamics
was found, too, in ``computer" games \cite{KimFractals2006},
wherein virtual players move on the screen 
following the inherent program of the computer software.

\section{Result 2: $H$-index evaluation}

To better understand this self-similarity,
we derived an averaged mean-square displacement of the time series $u_j(t)$ defined by
\begin{equation}
C_j(\tau) = \langle \left[ u_j(t+\tau) - u_j(t) \right]^2 \rangle.
\end{equation}
where the angle brackets $\langle \cdots \rangle$ indicate that an average is taken
over the entire time interval.
$C_j(\tau)$ describes the mean square fluctuation of the data profile
at time scale $\tau$.
If $C_j(\tau)$ follows a power law such as
\begin{equation}
C_j(\tau)\propto \tau^{2H} \quad (0<H<1),
\label{eq_02}
\end{equation}
then the corresponding process $u_j(t)$ is said to be fBm
with a Hurst exponent of $H$ \cite{BouchaudPhysRep1990}.
If $H>1/2$,
the inter-incremental correlation will be positive \cite{ShimaPRB2004},
indicating a strong trend in the time series marked by
a rather smooth data profile with a memory effect \cite{BouchaudPhysRep1990}.
Because the correlation between the increments disappears at a large time scale, 
in practice, eq.~(\ref{eq_02}) is valid only at $\tau < \eta_j$, where $\eta_j$ is
a characteristic time scale, in which case $C_j(\tau)$ for $\tau > \eta_j$ 
approaches a constant value.
The value of $\eta_j$, called the persistence time,
can be estimated from the cross-over point of $C_j(\tau)$
as shown below.

\begin{figure}
\resizebox{0.35\textwidth}{!}{%
  \includegraphics{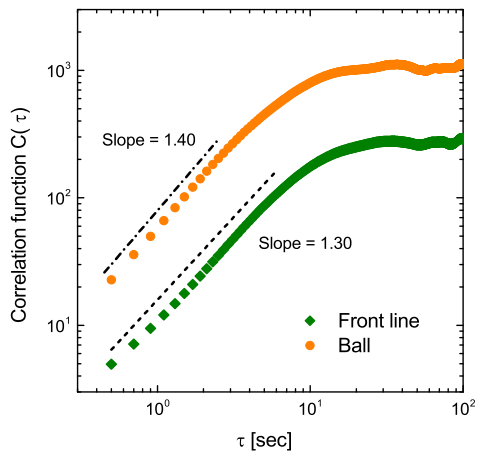}
}
\vspace*{5mm}
\caption{(color online)
Averaged mean-square displacement $C_j(\tau)$ for time lag $\tau$.
Power-law behaviours $C_j(\tau)\propto \tau^{2H}$ for exponents
$2H = 1.3$ and $1.4$ are highlighted with dashed and dashed-dotted lines, respectively.
A persistence time $\eta \sim 30$ s is
observed in both plots.}
\label{fig.5}
\end{figure}

Figure \ref{fig.5} shows calculated values of $C_j(\tau)$
for $u_j(t)$ [$j=$f, b] extracted from the Osaka-Adelaide data.
In the short term, the two curves of $C_j(\tau)$ obey power-law behaviours $C_j(\tau)\propto \tau^{2H}$
from 0.5 s to 5 s and to 2 s for the frontline and ball data, respectively, 
followed by convergence to constant values after 30 s;
these observations indicate a crossover time of $\eta \sim 30$ s.
The exponents for the frontline
and the ball dynamics were found to be $H = 0.65$ and $0.70$, respectively.
Based on this, we can conclude that football dynamics follow fBm behaviour
endowed with a memory effect.
This  conclusion was also  drawn from
the Urawa-Yokohama game data,
from which $H=0.66$ and $H=0.72$ 
were detected for the frontline and ball dynamics, respectively,
with the common persistence time $\eta= 30$ s.

\begin{figure}
\resizebox{0.5\textwidth}{!}{%
  \includegraphics{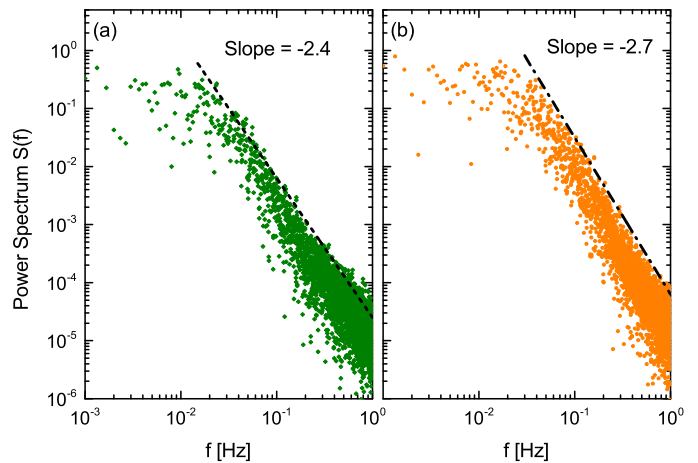}
}
\vspace*{5mm}
\caption{(color online)
Power spectrum $S(f)$ of the time variation in the frontline (a)
and  ball position (b).
The data at $f>f_c \sim 0.03$ s$^{-1}$
have been fitted to power laws of the form $S(f) \propto f^{-\alpha}$
with the exponents indicated.}
\label{fig.6}
\end{figure}

The observed power-law behaviour of $C_j(\tau)$ 
gives an important consequence of its Fourier power spectrum $S(f)$,
as the power spectrum of an fBm process is known to obey
the form \cite{GreisPRA1991}
\begin{equation}
S(f) \propto f^{-\alpha} \quad (\alpha = 2H+1)
\end{equation}
at $f>f_c\equiv \eta^{-1}$.
This power law is derived from the Wiener-Khintchin theorem
\begin{equation}
\Gamma(t) = \int S(f) e^{2\pi i f t} df
\end{equation}
with respect to the auto-correlation function
$\Gamma(\tau) = \langle u(t+\tau) u(t) \rangle - \langle u \rangle^2$,
which satisfies the relation
$C(\tau) = 2 \left[ \phi - \Gamma(\tau) \right]$,
where $\phi$ is the squared variance of the series $u(t)$ given by
$\phi = \left\langle \left[ u(t) - \langle u \rangle \right]^2 \right\rangle$.
Figure \ref{fig.6} demonstrates the validity of these arguments;
it shows the exponents $\alpha=2.4$ and $\alpha=2.7$ for the frontline and  ball dynamics,
respectively, obtained from the Osaka-Adelaide data.
The values of $\alpha$ obtained here are in fair agreement with
the theoretically predicted relation $\alpha = 2H + 1$.
The other game data (Urawa-Yokohama) yields $\alpha=2.5$ and $\alpha=2.8$ in the same order,
both of which satisfy the above relation.

\section{Result 3: Volatility in movement}

The observed persistence time less than 1 min
reflects a duration of ball possession
associated with the interactions of a small number of players.
Typically, the ball possession time for one team
does not exceed 1 min, but rather lasts only several tens of seconds at most;
after that, a ball interception or a dead ball
causes a turnover to the other team,
which resets the memory of the collective multiplayer motion.
In other words, the superiority of one team tends
to persist for 1 min or less,
following which the other team can have a chance to perform successfully.
Actually in 2002 FIFA World Cup$^{\rm TM}$, for instance,
89\% of all the ball possession events
were terminated in 30 s, 
within which 1 to 5 passes were made by one team \cite{Carling2005}.
Such the come-and-go, or turning of the tide,
which can occur on a minute-by-minute basis,
makes the dynamics of a football game unpredictable
and attractive to the audience.

\begin{figure}
\resizebox{0.48\textwidth}{!}{%
  \includegraphics{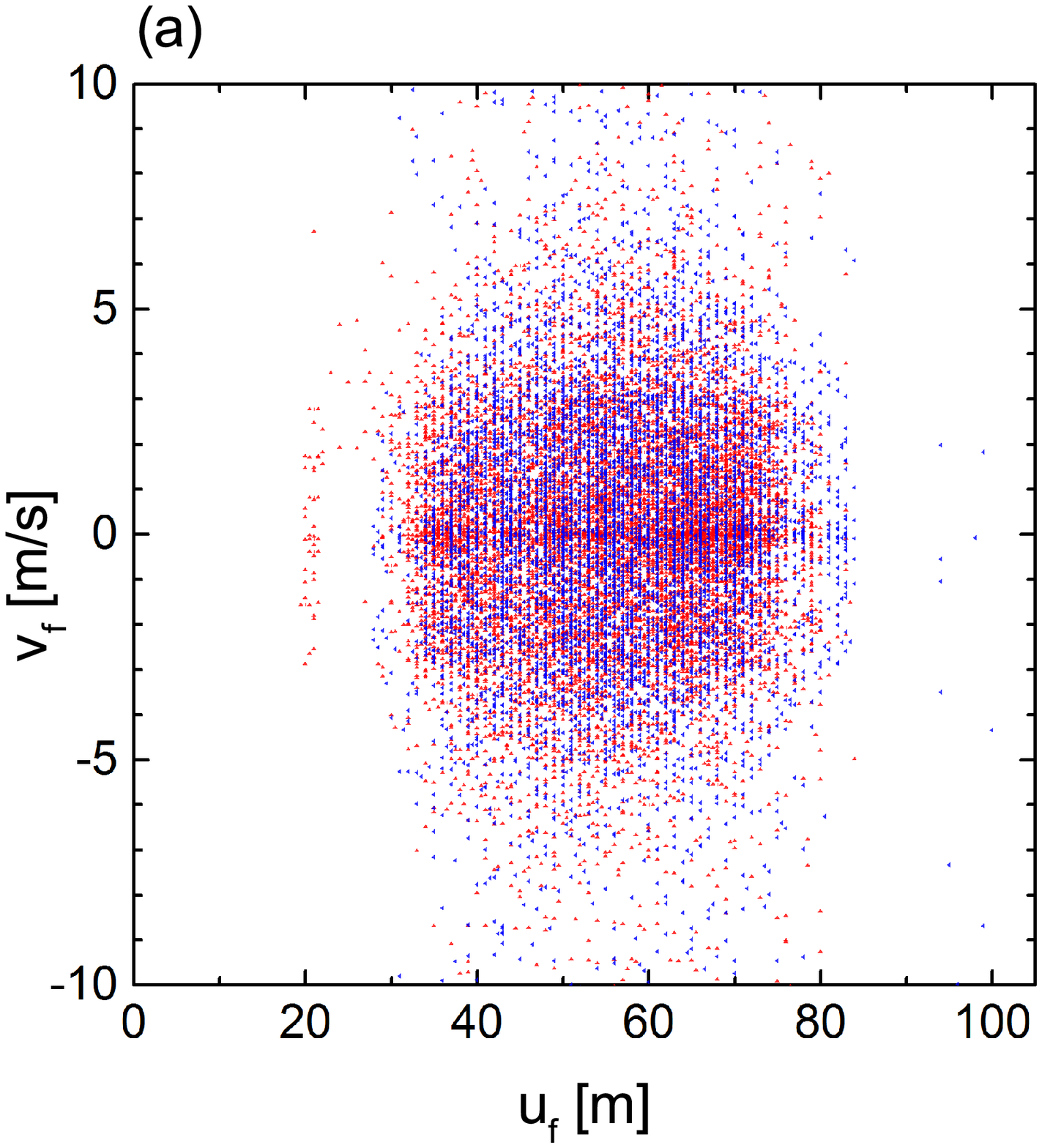}
  \includegraphics{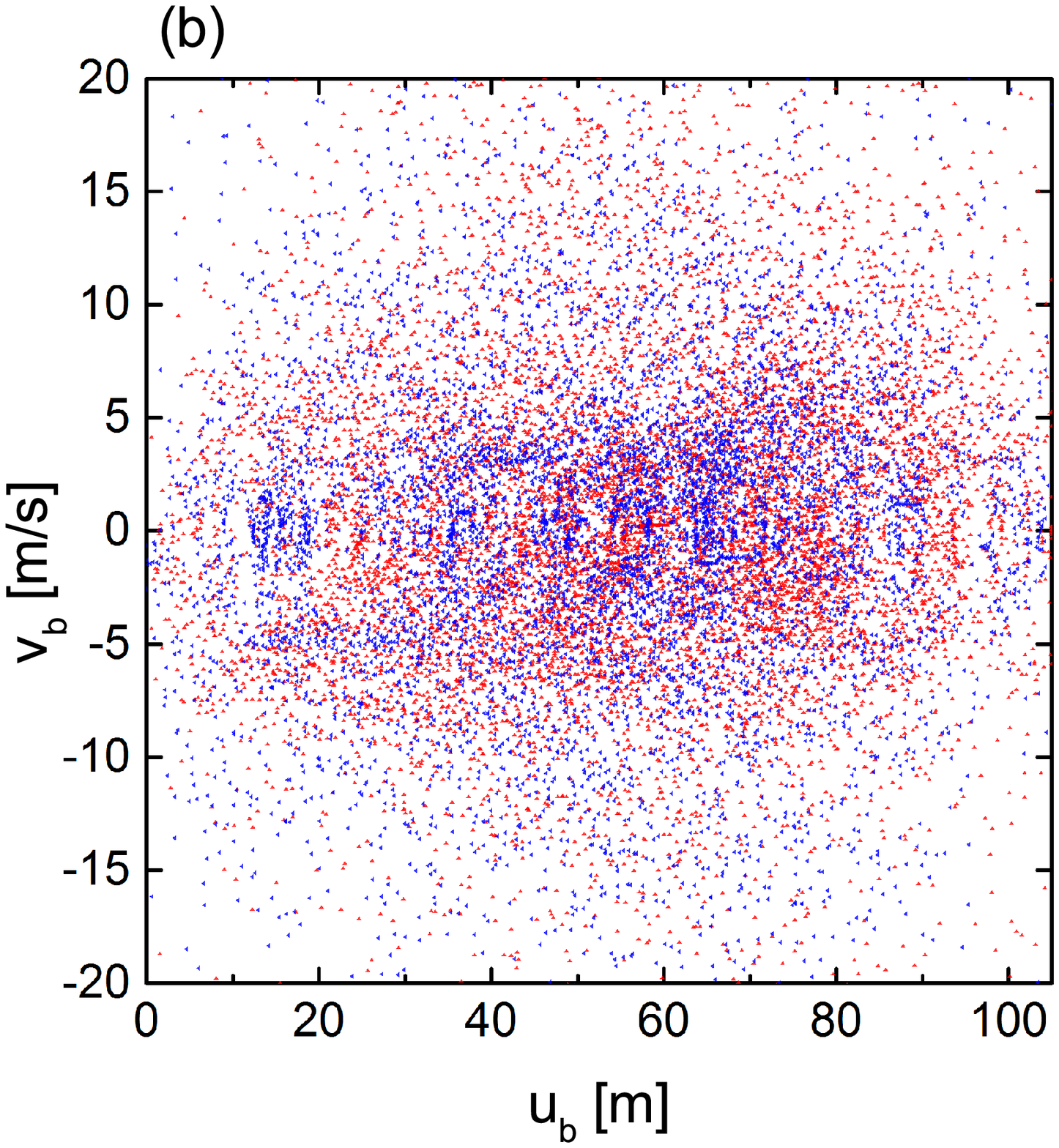}
}
\vspace*{5mm}
\caption{
(color online)
Distribution of the velocity of the front-line movement ($v_{\rm f}$) and that of the ball movement ($v_{\rm b}$). 
Triangles (coloured in red) indicate the velocity at the time when the ball is possessed by team Osaka 
and when the team's $u_j$ [j={\rm f,b}] has a value represented by the horizontal axis. 
Slanted triangles (coloured in blue) show the counterpart for team Adelaide.}
\label{fig.7}
\end{figure}

The  volatility in football dynamics can also be examined in a quantitative manner 
by considering the velocity of the front-line movement ($v_{\rm f}$) and that of the ball movement ($v_{\rm b}$). 
Figure \ref{fig.7} shows the distribution of $v_{\rm f}$ and $v_{\rm b}$; 
points indicate the velocity at the time when the ball is possessed by a team (Osaka or Adelaide) 
and when the team's $u_j$ $[j={\rm f,b}]$ is located at a position represented by the horizontal axis.
It should be emphasized that all the data points are distributed almost symmetrically with respect to the centre of the plot. 
This symmetric distribution evidences no relationship between the direction of the ball/front-line movement 
and team's ball possession, despite that ball possession may be apparently thought to bias the movement direction.
If the bias effect is dominant, then red (or blue) points should localize at the lower (upper) half in the plot, 
because the sign in $v_{\rm f}$ and $v_{\rm b}$ should be unaltered during the ball possession time.
However, such the data point localization does not occur in Fig.~\ref{fig.7}.
This fact means that, even when one team possesses the ball, neither $u_{\rm f}$ nor $u_{\rm b}$ 
tends to move monotonically toward the other side; 
instead, they involuntarily make a back pass and/or meandering dribble toward their own side.

\begin{figure}
\resizebox{0.45\textwidth}{!}{%
  \includegraphics{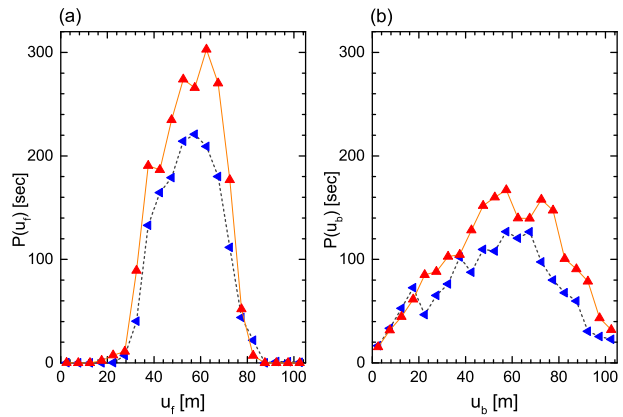}
}
\vspace*{5mm}
\caption{
(color online)
Histogram of the front-line position $(u_{\rm f})$ and the ball position $(u_{\rm b})$. 
Definitions of symbols, triangles (coloured in red), and slanted triangles
(coloured in blue) are the same as those used in Fig.~\ref{fig.7}.}
\label{fig.8}
\end{figure}

Figure \ref{fig.8} shows complementary proof to this volatility through histograms of $u_{\rm f}$ and $u_{\rm b}$ 
for the two teams during the total game duration. 
$P(u_j)$ $[j={\rm f,b}]$ along the vertical axis quantifies the duration (in seconds) 
within which each $u_j$ takes the corresponding position indicated on the horizontal axis. 
All the curves in the two histograms are symmetric with respect to the centre line of the pitch, 
implying the absence of spatial bias in the ball/player's movement. 
We thus conclude that the bias effect is insignificant in football dynamics, 
at least in considering the time evolution in $u_{\rm f}$ and $u_{\rm b}$.

To some readers, the absence of correlation between team's ball possession and 
the ball/front-line movement sounds unexpected, 
since one team having the ball try to bring it toward the opponent's goal.
Indeed, in many complex systems other than football dynamics, 
spatiotemporal fluctuations involve a certain class of biases \cite{PengPRE1994}, 
which may trigger crossovers in the scaling behavior of the systems \cite{HuPRE2001}.
We speculate that the disappearance of bias effects in football dynamics
stems from the turf domination scramble performed 
by two competing teams, both of which try to carry the ball toward their opponent's goal. 
Owing to the scramble, which describes the soccer game's momentum, players can 
hardly play the ball in their desired manner. 
Meanwhile, they move back-and-forth over the pitch in a manner quite different 
from a one-way manner to disconcert the opponent players. 
As a consequence of the competing gamesmanship, random fluctuations are enhanced 
and considerably exceed the bias effect.

\section{Concluding remark}

In conclusion, we have demonstrated through real-time game analyses that football dynamics follow fBm behaviour.
The longitudinal position of the team's frontline, $u_{\rm f}$,
and that of the ball, $u_{\rm b}$,
are both governed by the fBm mechanism,
resulting in a power-law behavior $C(\tau)\propto \tau^{2H}$
with Hurst exponent $H\sim 0.7$.
The persistence time $\eta$ has been estimated to be a few tens of seconds,
implying that the volatility of football dynamics from a long-term perspective
has a crucial role in making football games entertaining.
Perhaps a new `Fantasista' could emerge from the volatility of football dynamics.

%

\begin{acknowledgement}
We thank members of the DSA project for helpful suggestions.
We also thank to Masao Nakayama who provided expertise in football.
The work was supported by JSPS KAKENHI Grant Numbers 22650146, 23500711, 24240085,
and 25390147.
\end{acknowledgement}


\end{document}